\documentclass[a4paper]{article}

\usepackage[margin=1in]{geometry}

\usepackage[T1]{fontenc}
\newcommand{\changefont}[3]{
\fontfamily{#1} \fontseries{#2} \fontshape{#3} \selectfont}

\changefont{ptm}{m}{n}

\usepackage{setspace} 
\usepackage{graphicx}    

\usepackage{amsfonts,amssymb,amsmath}
\usepackage{graphics}
\usepackage{mathrsfs}
\usepackage{color}

\newtheorem{theorem}{Theorem}[section]

\newtheorem{corollary}{Corollary}[section]
\newtheorem{lemma}{Lemma}[section]

\newtheorem{definition}{Definition}[section]

\long\def\symbolfootnote[#1]#2{\begingroup%
\def\thefootnote{\fnsymbol{footnote}}\footnote[#1]{#2}\endgroup} 

\begin{document}

\begin{center}
\Large \textbf{Replication of Period-Doubling Route to Chaos in Coupled Systems with Delay}
\end{center}

\begin{center}
\normalsize \textbf{Mehmet Onur Fen$^{1,}$\symbolfootnote[1]{Corresponding Author. E-mail: monur.fen@gmail.com, Tel: +90 312 585 0217}, Fatma Tokmak Fen$^2$} \\
\vspace{0.2cm}
\textit{\textbf{$^1$Department of Mathematics, TED University, 06420 Ankara, Turkey}}

\vspace{0.1cm}
\textit{\textbf{$^2$Department of Mathematics, Gazi University, 06500 Ankara, Turkey}}
\vspace{0.1cm}
\end{center}

\vspace{0.3cm}

\begin{center}
\textbf{Abstract}   
\end{center}

\vspace{-0.2cm}

\noindent\ignorespaces

In this study, replication of a period-doubling cascade in coupled systems with delay is rigorously proved under certain assumptions, which guarantee the existence of bounded solutions and replication of sensitivity. A novel definition for replication of sensitivity is utilized, in which the proximity of solutions is considered in an interval instead of a single point. Examples with simulations supporting the theoretical results concerning sensitivity and period-doubling cascade are provided.

\vspace{0.2cm}
 
\noindent\ignorespaces \textbf{Keywords:} Replication of chaos; Period-doubling cascade; Sensitivity; Systems with delay; Unidirectional coupling

\vspace{0.2cm}

\noindent\ignorespaces \textbf{2010 Mathematics Subject Classification:} {34K23; 34K18; 37D45; 34K13}


\section{Introduction} \label{intro}

The occurrence of time delays is an issue that is crucial for nonlinear processes. In the case that time delays are taken into consideration, delay differential equations can be utilized in modeling of such processes. Infinite-dimensional dynamical systems can arise from delay differential equations \cite{Hale71,Smith11}, and such equations can exhibit chaotic behavior, even in scalar case \cite{Walther81,Heiden83}. Delay differential equations are useful in various fields such as neural networks, secure communication, robotics, economics, and lasers \cite{Lakshmanan18}-\cite{Erneux19}. Some open problems concerning differential equations with delay can be found in the papers \cite{Walther14}-\cite{Diblik19}. Another phenomenon that can be observed in dynamics of nonlinear systems is period-doubling cascade, which is capable of giving rise to the emergence of chaos. Likewise systems with delay, the presence of period-doubling cascades as well as chaos can be observed and have applications in a variety of scientific areas \cite{Volos20}-\cite{Ndofor18}.

We understand chaos as the presence of sensitivity, which can be considered as the main ingredient of chaos \cite{Lorenz63}-\cite{Robinson95}, and infinitely many unstable periodic solutions in a compact region. 
Motivated by the effective scientific roles of chaos and systems with delay, in the present study, we consider the replication of period-doubling route to chaos in unidirectionally coupled systems in which the secondary system is with delay. More precisely, we take into account the systems
\begin{eqnarray} \label{system1}
x'(t)=F(t,x(t))
\end{eqnarray}
and
\begin{eqnarray} \label{system2}
y'(t)=Ay(t)+G(t,x(t),y(t-\tau))
\end{eqnarray}
where the functions $F:\mathbb R \times \mathbb R^m  \to \mathbb R^m$ and $G:\mathbb R \times \mathbb R^m  \times \mathbb R^n \to \mathbb R^n$ are continuous in all of their arguments, all eigenvalues of the matrix $A\in\mathbb R^{n \times n}$ have negative real parts, and $\tau$ is a positive number. Our purpose is to rigorously prove that system (\ref{system2}) exhibits chaotic motions, provided that the same is true for system (\ref{system1}) under certain conditions that will be mentioned in the next section. These assumptions are required in order to verify the existence of bounded solutions and replication of sensitivity.

The existence of chaos in delay differential equations was demonstrated in the studies \cite{Walther81,Heiden83,Wayda96}. The results of the papers \cite{Walther81,Heiden83} are based on the Li-Yorke definition of chaos \cite{Li75,Marotto78}. Utilizing the results obtained in their study \cite{Wayda95}, Lani-Wayda and Walther \cite{Wayda96} constructed a delay differential equation possessing chaotic behavior. On the other hand, global attractors for delay differential equations were investigated in \cite{Krisztin01b,Krisztin01a}. The global attractiveness of a three-dimensional compact invariant set of a differential equation modeling a system governed by delayed positive feedback and instantaneous damping was shown by Krisztin and Walther \cite{Krisztin01b}. In addition to the presence of a global attractor in the dynamics of a differential equation with state-dependent delay, one of its topological properties were investigated by \cite{Krisztin01a}. Moreover, results concerning periodic solutions of state-dependent delay differential equations were presented by Kuang and Smith \cite{Kuang92}. The technique provided in this study is different compared to the papers \cite{Walther81,Heiden83,Wayda96} such that we take into account replication of sensitivity and period-doubling cascades in unidirectionally coupled systems.

Regular inputs such as periodic, quasi-periodic, and almost periodic motions can lead to the formation of outputs of the same types in dynamics of certain types of differential equations \cite{Corduneanu68,Fink74}. The main idea of our investigation is the usage of chaotic motions as inputs in systems with delay, and it is demonstrated that chaotic outputs are obtained. The inputs are supplied from solutions of another system possessing chaos. The reader is referred to the papers \cite{Enciso06,Scardovi10} for some applications of input-output systems.

The foundations of chaos generation in systems of differential equations by means of perturbations and impulsive actions were laid by Akhmet \cite{Akhmet1}-\cite{Akhmet3}. An answer to the question whether continuous chaotic inputs generate chaotic outputs was given in the study \cite{Fen13} for systems without delay. It was rigorously proved in \cite{Fen13} that under certain conditions chaotic dynamics of a system of differential equations can be replicated by another system under unidirectional coupling between them. Chaos in the senses of Devaney \cite{Devaney87} and Li-Yorke \cite{Li75} as well as period-doubling cascade were considered by Akhmet and Fen \cite{Fen13}. Moreover, the study \cite{Fen2019} was concerned with replication of period-doubling route to chaos in systems with impulsive actions. Systems with delay were not taken into account in the studies \cite{Fen13,Fen2019}. The main novelty of the present research is the consideration of replication of chaos problem for systems with delay. Due to the presence of delay, a novel definition for replication of sensitivity is provided and the contraction mapping principle is utilized for its verification. The obtained results are valid for systems with arbitrary high dimensions. The book \cite{Akhmet16} comprises some applications of the replication of chaos technique to neural networks, economics, and weather dynamics.

In the literature, unidirectionally coupled chaotic systems have been considered within the scope of synchronization \cite{Pecora90}-\cite{Gon04}. In the case of identical systems, synchronization occurs when asymptotic proximity of the states of the drive and response systems is valid \cite{Pecora90}. For the presence of synchronization in the dynamics of non-identical systems, the asymptotic proximity is considered with the help of a functional relation, which determines the phase space trajectory of the response system from the trajectory of the drive \cite{Rulkov95}. The approach utilized in this study is different from synchronization of chaos since the coupled system (\ref{system1})-(\ref{system2}) is not taken into account from the asymptotic point of view. For that reason, following the terminology of paper \cite{Fen13}, we call system (\ref{system1}) the \textit{generator} and system (\ref{system2}) the \textit{replicator}.

The rest of the paper is organized as follows. In the next section, preliminary results and conditions on the coupled system (\ref{system1})-(\ref{system2}), which are required for replication of sensitivity and the existence of unstable periodic solutions, are provided. In Section \ref{sensitivitysection}, replication of sensitivity is theoretically investigated. Section \ref{pdcsection}, on the other hand, is concerned with replication of period-doubling cascade. Section \ref{secexamples} is devoted to examples in which the Lorenz system \cite{Lorenz63} and Duffing equation \cite{Sato83} are utilized as generator systems. Finally, some concluding remarks are given in Section \ref{secconc}.

\section{Preliminaries} \label{prelim}

Our main assumption on the generator system (\ref{system1}) is the existence of a nonempty set $\mathscr{A}$ of all solutions of the system that are uniformly bounded on $\mathbb R$. In this case there exists a compact set $\Lambda \subset \mathbb R^m$ such that the trajectories of all solutions that belong to $\mathscr{A}$ lie inside $\Lambda$. We also assume that there exists a positive number $T$ such that $$F(t+T,x)=F(t,x)$$ and 
\begin{eqnarray} \label{periodofG}
\ \ \ \ \ \  G(t+T,x,y)=G(t,x,y)
\end{eqnarray}
for all $t\in\mathbb R$, $x \in \mathbb R^m$, and $y \in \mathbb R^n$.

Since we suppose that all eigenvalues of the matrix $A$ in the replicator system (\ref{system2}) have negative real parts, there exist numbers $K\geq 1$ and $\omega >0$ such that $\left\|e^{At}\right\|\leq K e^{-\omega t}$ for all $t \geq 0$. 

Throughout the paper, we make use of the usual Euclidean norm for vectors and the spectral norm for square matrices.

The following conditions are required.
\begin{enumerate}
\item[\bf (C1)] There exists a positive number $L_F$ such that $\displaystyle  \left\| F(t,x)-F(t,\widetilde{x}) \right\| \leq L_F\left\|x-\widetilde{x}\right\|$ for all $t \in \mathbb R$ and $x,\widetilde{x}\in \Lambda$; 
\item[\bf (C2)] There exists a positive number $L_1$ such that $\left\|G(t,x,y)-G(t,\widetilde{x},y)\right\| \geq L_1 \left\|x-\widetilde{x}\right\|$ for all $t\in \mathbb R$, $x,\widetilde{x}\in \Lambda$, and $y\in\mathbb R^n$;
\item[\bf (C3)] There exists a positive number $L_2$ such that $\left\|G(t,x,y)-G(t,\widetilde{x},y)\right\|\leq L_2 \left\|x-\widetilde{x}\right\|$ for all $t\in \mathbb R$, $x,\widetilde{x}\in \Lambda$, and $y\in\mathbb R^n$;
\item[\bf (C4)] There exists a positive number $L_3$ such that
$\left\|G(t,x,y)-G(t,x,\widetilde{y})\right\|\leq L_3\left\|y-\widetilde{y}\right\|$ for all $t\in \mathbb R$, $x \in \Lambda$, and $y, \widetilde{y} \in\mathbb R^n$;
\item[\bf (C5)] There exists a positive number $M_G$ such that $\displaystyle \sup_{t\in \mathbb R, x \in \Lambda, y\in\mathbb R^n}\left\|G(t,x,y)\right\|\leq M_G$;
\item[\bf (C6)] $\omega-2KL_3e^{\omega \tau/2}>0$.
\end{enumerate}

For the existence and uniqueness of the bounded solutions of system (\ref{system2}), the conditions $(C4)$ and $(C5)$ are utilized. Conditions $(C1)$, $(C3)$, $(C4)$, and $(C6)$, on the other hand, are required to show the proximity of the bounded solutions of (\ref{system2}) on a closed interval with length $\tau$ in the verification of replication of sensitivity. Moreover, the condition $(C2)$ is used in the replication of sensitivity to show the divergence of the bounded solutions, and the condition $(C6)$ is required also in their global exponential stability.

Suppose that the conditions $(C4)$, $(C5)$ hold. For a fixed solution $x \in \mathscr{A}$ of system (\ref{system1}), it can be verified that a function $y(t)$ which is bounded on the whole real axis is a solution of system (\ref{system2}) if and only if the integral equation
$$y(t)=\displaystyle \int \limits_{-\infty}^{t} e^{A(t-s)} G(s,x(s),y(s-\tau)) ds$$
is satisfied.
Denote by $\mathscr{C}$ the set of continuous functions $\varphi:\mathbb R \to \mathbb R^n$ with $\left\|\varphi\right\|_{\infty} \leq M_0$, where $$\left\|\varphi\right\|_{\infty}=\displaystyle \sup_{t \in \mathbb R} \left\|\varphi(t)\right\|$$ and 
\begin{eqnarray} \label{numberm0}
M_0=\displaystyle \frac{K M_{G}}{\omega}.
\end{eqnarray}
Let us define the operator $\Gamma$ on $\mathscr{C}$ through the equation
$$\Gamma \varphi(t)= \displaystyle \int \limits_{-\infty}^{t} e^{A(t-s)} G(s,x(s), \varphi(s-\tau)) ds.$$
If $\varphi$ is a function in $\mathscr{C}$, then one can confirm that $\left\|\Gamma \varphi\right\|_{\infty} \leq M_0$, which yields $\Gamma (\mathscr{C}) \subseteq \mathscr{C}$. 
Additionally, if $\varphi_1$, $\varphi_2$ belong to $\mathscr{C}$, then $\left\|\Gamma \varphi_1 - \Gamma \varphi_2\right\|_{\infty} \leq \displaystyle \frac{K L_{3}}{\omega} \left\|\varphi_1-\varphi_2\right\|_{\infty}$. Therefore, if $\omega- K L_3 >0$, then the operator $\Gamma$ is a contraction. For that reason, if the conditions $(C4)$, $(C5)$ hold and the inequality $\omega-KL_3>0$ is valid, then for each fixed solution $x \in \mathscr{A}$ of system (\ref{system1}), there exists a unique solution $\phi_{x}$ of system (\ref{system2}) which is bounded on the whole real axis such that $$\displaystyle \sup_{t \in \mathbb R}\left\|\phi_{x}(t)\right\|\leq M_0$$ and  
\begin{eqnarray} \label{bddsolnrelation}
\phi_{x}(t)=\displaystyle \int\limits_{-\infty}^{t} e^{A(t-s)} G(s,x(s),\phi_{x}(s-\tau) ) ds.
\end{eqnarray}

It is worth noting that the condition $(C6)$ implies the inequality $\omega-KL_3>0$. 
Moreover, if conditions $(C4)-(C6)$ are satisfied, then for each $x \in \mathscr{A}$ the bounded solution $\phi_{x}(t)$ of system (\ref{system2}) is globally exponentially stable \cite{Driver77}.

To investigate the replication of sensitivity theoretically, we introduce the set of uniformly bounded functions
\begin{eqnarray} \label{setB}
\mathscr{B}=\left\{\phi_{x}(t) :~ x\in\mathscr{A}\right\}.
\end{eqnarray} 
There is a one-to-one correspondence between the sets $\mathscr{A}$ and $\mathscr{B}$ under the condition $(C2)$. In other words, for each solution $x \in \mathscr{A}$ of generator (\ref{system1}) there exists a unique bounded solution $\phi_{x} \in \mathscr{B}$ of replicator (\ref{system2}), and vice versa.

\section{Replication of Sensitivity} \label{sensitivitysection} 
 
The definition of sensitivity for system (\ref{system1}) is as follows. 
 
\begin{definition} \label{sensitivtydefn1} \cite{Fen13}.  
System (\ref{system1}) is called sensitive if there exist positive numbers $\epsilon_0$ and $\Delta$ such that for an arbitrary positive number $\delta_0$ and for each $x \in\mathscr{A}$, there exist $\overline{x} \in \mathscr{A}$, $t_0\in\mathbb R$, and an interval $J\subset [t_0,\infty)$ with a length no less than $\Delta$ such that $\left\|x(t_0)-\overline{x}(t_0)\right\|<\delta_0$ and $\left\|x(t)-\overline{x}(t)\right\|>\epsilon_0$ for all $t\in J$.
\end{definition}

The next definition is concerned with the replication of sensitivity by systems with delay.

\begin{definition} \label{sensitivtydefn2}
System (\ref{system2}) replicates the sensitivity of system (\ref{system1}) if there exist positive numbers $\epsilon_1$ and $\overline{\Delta}$ such that for an arbitrary positive number $\delta_1$ and for each bounded solution $\phi_{x} \in \mathscr{B}$, there exist a bounded solution $\phi_{\overline{x}} \in \mathscr{B}$, $t_0 \in \mathbb R$, and an interval $\widetilde{J}\subset [t_0,\infty)$ with a length no less than $\overline{\Delta}$ such that $\displaystyle \sup_{t\in [t_0-\tau, t_0]}\left\|\phi_{x}(t)-\phi_{\overline{x}}(t)\right\|<\delta_1$ and $\left\|\phi_{x}(t)-\phi_{\overline{x}}(t)\right\|>\epsilon_1$ for all $t \in \widetilde{J}$.
\end{definition}

The main result of the present section is provided in the next theorem.
In the proof of the theorem, first of all, utilizing the initial proximity of two solutions of system (\ref{system1}) in $\mathscr{A}$ we estimate the distance between them backward in time. Then, this estimation is used to verify the initial proximity of the bounded solution of system (\ref{system2}) on an interval of length $\tau$ by means of the contraction mapping principle. Finally, an equicontinuous family of functions is constructed based on the equicontinuity of both $\mathscr{A}$ and the bounded solutions of system (\ref{system2}), and it is use to show the divergence of the bounded solutions of (\ref{system2}). The provided proof technique makes it possible to determine the numbers $\epsilon_1$ and $\overline{\Delta}$ mentioned in Definition \ref{sensitivtydefn2}.

\begin{theorem} \label{delaymaintheorem1}
Assume that the conditions $(C1)-(C6)$ are valid. If system (\ref{system1}) is sensitive, then system (\ref{system2}) replicates the sensitivity of (\ref{system1}).
\end{theorem}

\noindent \textbf{Proof.}
Fix an arbitrary positive number $\delta_1$ and a bounded solution $\phi_{x} \in \mathscr{B}$ of system (\ref{system2}). Let us denote 
\begin{eqnarray} \label{eqnr1}
R_1=\displaystyle \frac{2KM_0 \omega}{\omega-2KL_3e^{\omega \tau/2}},
\end{eqnarray}
where $M_0$ is the number defined by (\ref{numberm0}), and
\begin{eqnarray} \label{eqnr2}
R_2=\displaystyle \frac{KL_2}{\omega-KL_3}. 
\end{eqnarray}
The numbers $R_1$ and $R_2$  are positive by condition $(C6)$. Suppose that $\delta_0=\gamma \delta_1 e^{-NL_F}$, where $\gamma$ is a positive number such that  
\begin{eqnarray} \label{ineqgamma}
\gamma < \displaystyle \frac{1}{R_1+R_2}
\end{eqnarray}
and $N$ is a positive number satisfying the inequality 
\begin{eqnarray} \label{ineqnumbern}
N \geq \tau + \displaystyle \frac{2}{\omega} \ln\left(\frac{1}{\gamma \delta_1}\right).
\end{eqnarray}  

Since system (\ref{system1}) is sensitive, there exist positive numbers $\epsilon_0$ and $\Delta$ such that 
\begin{eqnarray} \label{ineqdelta0}
\left\|x(t_0)-\overline{x}(t_0)\right\|< \delta_0
\end{eqnarray} 
and 
\begin{eqnarray} \label{ineqeps0}
\left\|x(t)-\overline{x}(t)\right\|>\epsilon_0, ~t \in J, 
\end{eqnarray} 
for some $\overline{x} \in \mathscr{A}$, $t_0\in\mathbb R$, and for some interval $J \subset [t_0,\infty)$ with a length no less than $\Delta$.

Making use of the relation
\begin{eqnarray*}
x(t)-\overline{x}(t) = x(t_0)-\overline{x}(t_0) + \displaystyle \int\limits_{t_0}^{t} \left( F(s,x(s))- F(s, \overline{x}(s)) \right) ds,
\end{eqnarray*}
it can be verified for $t \in [t_0-N,t_0]$ that
\begin{eqnarray*}
\left\|x(t)-\overline{x}(t) \right\|\leq \left\|x(t_0)-\overline{x}(t_0)\right\| + \bigg| \displaystyle \int\limits_{t_0}^{t} L_F\left\|x(s)-\overline{x}(s) \right\|ds\bigg|,
\end{eqnarray*}
Applying the Gronwall-Bellman inequality \cite{Corduneanu08} we obtain that
\begin{eqnarray*}
\left\|x(t)-\overline{x}(t) \right\|\leq \left\|x(t_0)-\overline{x}(t_0)\right\|e^{L_F \left|t-t_0\right|}, ~ t\in [t_0-N,t_0].
\end{eqnarray*}
Therefore, $\left\|x(t)-\overline{x}(t) \right\| < \gamma \delta_1$ for $t \in [t_0-N,t_0]$ in accordance with the inequality (\ref{ineqdelta0}).

One can confirm that the function $\psi(t)=\phi_{x}(t)- \phi_{\overline{x}}(t)$ is a solution of the system
\begin{eqnarray}  \label{psieqn}
\psi'(t)=A \psi(t)+G(t,x(t),\psi(t-\tau)+\phi_{\overline{x}}(t-\tau))-G(t,\overline{x}(t),\phi_{\overline{x}}(t-\tau)).
\end{eqnarray}
Accordingly, the equation
\begin{eqnarray*} 
\psi(t) &=&  e^{A(t-t_0+N)} \left(  \phi_{x}(t_0-N)- \phi_{\overline{x}}(t_0-N) \right)  \\
&& + \displaystyle \int\limits_{t_0-N}^{t} e^{A(t-s)} \left( G(s,x(s),\psi(s-\tau)+\phi_{\overline{x}}(s-\tau) ) - G(s,\overline{x}(s),\phi_{\overline{x}}(s-\tau))  \right) ds
\end{eqnarray*}
is satisfied for $t\geq t_0-N$.

Let us denote by $\mathscr{H}$ the set of continuous functions $\psi(t)$ defined on $\mathbb R$ such that
\begin{eqnarray} \label{ineqpsi1}
\left\|\psi(t)\right\| \leq R_1 e^{-\omega(t-t_0+N)/2} + R_2\gamma\delta_1
\end{eqnarray}
for $t_0-N-\tau \leq t \leq t_0$, where the numbers $R_1$ and $R_2$ are respectively defined by the equations (\ref{eqnr1}) and (\ref{eqnr2}), and
$
\displaystyle \left\|\psi\right\|_{\infty} \leq 2K\left(M_0+\frac{M_G}{\omega}\right)
$
in which $\left\|\psi\right\|_{\infty} = \displaystyle \sup_{t\in\mathbb R} \left\|\psi(t)\right\|.$ Define an operator $\Pi$ on $\mathscr{H}$ through the equation
\begin{eqnarray*} 
\Pi \psi(t)= \left\{\begin{array}{ll}   \phi_{x}(t) - \phi_{\overline{x}}(t),  ~ t < t_0-N, \\ \\
 e^{A (t-t_0+N)} \left( \phi_{x}(t_0-N) - \phi_{\overline{x}}(t_0-N)  \right)  \\
 +\displaystyle \int\limits_{t_0-N}^{t} e^{A(t-s)}  \left( G(s,x(s),\psi(s-\tau)+\phi_{\overline{x}}(s-\tau) ) - 
 G(s,\overline{x}(s),\phi_{\overline{x}}(s-\tau) ) \right)  ds, ~ t\geq t_0-N. 
\end{array}\right.
\end{eqnarray*}

First of all, we will show that $\Pi \left(\mathscr{H}\right) \subseteq \mathscr{H}$. Suppose that $\psi(t)$ belongs to $\mathscr{H}$. If $t_0-N \leq t \leq t_0$, then it can be obtained using inequality (\ref{ineqpsi1}) that 
\begin{eqnarray*}
\left\|\Pi \psi(t)\right\| & \leq & \left\| e^{A(t-t_0+N)} \right\| \left\|\phi_{x}(t_0-N) - \phi_{\overline{x}}(t_0-N)\right\| \\
&& + \displaystyle \int\limits_{t_0-N}^{t} \left\|e^{A(t-s)}\right\|  \left\|G(s,x(s),\psi(s-\tau)+\phi_{\overline{x}}(s-\tau) ) - 
 G(s,x(s),\phi_{\overline{x}}(s-\tau) )\right\| ds \\
 && + \displaystyle \int\limits_{t_0-N}^{t} \left\|e^{A(t-s)}\right\|  
 \left\|G(s,x(s),\phi_{\overline{x}}(s-\tau) ) - 
 G(s,\overline{x}(s),\phi_{\overline{x}}(s-\tau) )\right\| ds \\
& \leq & 2KM_{0} e^{-\omega(t-t_0+N)} + \displaystyle \int\limits_{t_0-N}^{t} KL_3  e^{-\omega(t-s)} \left(R_1 e^{-\omega(s-\tau-t_0+N)/2} +R_2 \gamma \delta_1 \right) ds \\
&& +  \displaystyle \int\limits_{t_0-N}^{t} K L_{2} \gamma \delta_1 e^{-\omega(t-s)} ds \\
&<& \displaystyle 2K \left(M_0 + \frac{L_3 R_1 e^{\omega \tau/2}}{\omega}  \right) e^{-\omega(t-t_0+N)/2} 
+ \frac{K\gamma \delta_{1}}{\omega} \left(L_{2} + L_{3} R_{2} \right) \\
&=& R_1 e^{-\omega(t-t_0+N)/2} + R_2 \gamma \delta_1,
\end{eqnarray*}
since equations (\ref{eqnr1}) and (\ref{eqnr2}) respectively imply that $$\displaystyle 2K \left(M_0 + \frac{L_3 R_1 e^{\omega \tau/2}}{\omega}  \right) =R_1$$ and $$\displaystyle \frac{K}{\omega} \left(L_{2} + L_{3} R_{2} \right)=R_2.$$

Additionally, since $2M_0 < R_1$, the inequality $\left\|\Pi \psi(t)\right\|<R_1 e^{-\omega(t-t_0+N)/2} + R_2 \gamma \delta_1$ is also valid for $t_0-N-\tau \leq t < t_0-N$. On the other hand, it can be confirmed that $\displaystyle \left\|\Pi \psi\right\|_{\infty} \leq 2K\left(M_0+\frac{M_G}{\omega}\right)$. Thus, $\Pi \left(\mathscr{H}\right) \subseteq \mathscr{H}$.
 
Now, our purpose is to verify that the operator $\Pi$ is a contraction. Let $\psi_{1}(t)$ and $\psi_{2}(t)$ be functions that belong to $\mathscr{H}$. The inequality
\begin{eqnarray*}
 \left\|\Pi \psi_1(t)-\Pi \psi_2(t)\right\| 
  & \leq & \displaystyle \int\limits_{t_0-N}^{t} \left\|e^{A(t-s)}\right\| 
\big\| G(s,x(s),\psi_1(s-\tau)+\phi_{\overline{x}}(s-\tau)) \\
&& ~~~~~~~~~~~~~~~~~~   -G(s,x(s),\psi_2(s-\tau)+\phi_{\overline{x}}(s-\tau))  \big\| ds \\
& < & \frac{KL_3}{\omega} \left(1-e^{-\omega (t-t_0+N)}\right) \left\|\psi_1-\psi_2\right\|_{\infty}
\end{eqnarray*}
is valid for $t \geq t_0-N$. 
Moreover, $\left\|\Pi \psi_1(t)-\Pi \psi_2(t)\right\| =0$ for $t<t_0-N$. Hence, $$\displaystyle \left\|\Pi \psi_1 -\Pi \psi_2 \right\|_{\infty}\leq \frac{KL_3}{\omega} \left\|\psi_1-\psi_2\right\|_{\infty},$$ and the operator $\Pi$ is a contraction since the inequality $\displaystyle \frac{KL_3}{\omega}<1$ holds by condition $(C6)$.

According to the uniqueness of solutions of system (\ref{psieqn}), the function $\psi(t)=\phi_{x}(t)- \phi_{\overline{x}}(t)$ is the unique fixed point of the operator $\Pi.$ 
Let us denote 
\begin{eqnarray*}
\psi_0(t)= \left\{\begin{array}{ll}   \phi_{x}(t) - \phi_{\overline{x}}(t),  ~ t < t_0-N, \\ \\
e^{A (t-t_0+N)} \left( \phi_{x}(t_0-N) - \phi_{\overline{x}}(t_0-N)  \right) , ~ t \geq t_0-N,
\end{array}\right.
\end{eqnarray*}
which belongs to $\mathscr{H}.$ The sequence of functions $\left\{\psi_k(t)\right\},$ where $\psi_{k+1}(t) = \Pi \psi_k(t),$ $k =0,1,2,\ldots,$ converges to $\phi_{x}(t)- \phi_{\overline{x}}(t)$ on $\mathbb R.$
Thus, $$\left\|\phi_{x}(t)- \phi_{\overline{x}}(t)\right\| \leq R_1 e^{-\omega(t-t_0+N)/2} + R_2\gamma\delta_1$$ for $t_0-N-\tau \leq t \leq t_0$. Using the inequalities (\ref{ineqgamma}) and (\ref{ineqnumbern}), one can confirm for $t_0-\tau \leq t \leq t_0$ that
\begin{eqnarray*}
\left\|\phi_{x}(t)- \phi_{\overline{x}}(t)\right\| \leq R_1 e^{-\omega(-\tau+N)/2} + R_2\gamma\delta_1 \leq (R_1+R_2) \gamma \delta_1<\delta_1.
\end{eqnarray*}
Hence, we have  $\displaystyle \sup_{t\in [t_0-\tau, t_0]}\left\|\phi_{x}(t)-\phi_{\overline{x}}(t)\right\|<\delta_1$.

Next, we will show the existence of positive numbers $\epsilon_1$ and $\overline{\Delta}$ such that $\left\|\phi_{x}(t)- \phi_{\overline{x}}(t)\right\|>\epsilon_1$ for all $t\in\widetilde{J}$, where $\widetilde{J} \subset [t_0,\infty)$ is an interval with length $\overline{\Delta}$.

Let $M_F=\displaystyle \sup_{t\in\mathbb R, x \in \Lambda} \left\|F(t,x)\right\|$. Both of the sets $\mathscr{A}$ and 
$\mathscr{B}_0=\left\{ \phi_{x}(t-\tau): ~ x \in \mathscr{A} \right\}$ are equicontinuous families on $\mathbb R$ since $\displaystyle \sup_{t\in\mathbb R} \left\|x'(t)\right\| \leq M_F$ and $\displaystyle \sup_{t\in\mathbb R} \left\|\phi'_{x}(t-\tau)\right\|\leq \left\|A\right\|M_0+M_G$ for each solution $x\in\mathscr{A}$ of system (\ref{system1}).

Suppose that $G(t,x,y)=\left(G_1(t,x,y),G_2(t,x,y),\ldots,G_n(t,x,y) \right)$, where $G_i(t,x,y)$, $i=1,2,\ldots,n$, are real valued functions. Let us denote $\Lambda_0=\left\{ y \in \mathbb R^n: ~\left\|y\right\| \leq M_0 \right\}$ and define the function $\overline{G}: \mathbb R \times \Lambda \times \Lambda \times \Lambda_0 \to \mathbb R^n$ by $\overline{G}(t,x_1,x_2,y)=G(t,x_1,y)-G(t,x_2,y)$. Due to the periodicity of the function $G(t,x,y)$ in $t$, the function $\overline{G}(t,x_1,x_2,y)$ is uniformly continuous on $\mathbb R \times \Lambda \times \Lambda \times \Lambda_0$. Therefore, the set of functions
$$
\mathscr{F}=\left\{G_i(t,x(t),\phi_{x}(t-\tau))- G_i(t,\overline{x}(t),\phi_{x}(t-\tau)): ~ 1\leq i \leq n, ~ x, \overline{x} \in \mathscr{A} \right\}
$$
is an equicontinuous family on $\mathbb R$. Thus, there exists a positive number $\xi< \Delta$, which is independent of the functions $x(t)$ and $\overline{x}(t)$, such that for each $t_1,t_2 \in \mathbb R$ with $\left|t_1-t_2\right|<\xi$, the inequality 
\begin{eqnarray} \label{proofineq1}
 &&\big| \left(  G_i(t_1,x(t_1),\phi_{x}(t_1-\tau)) - G_i(t_1,\overline{x}(t_1),\phi_{x}(t_1-\tau))   \right) \nonumber \\ 
&& - \left(  G_i(t_2,x(t_2),\phi_{x}(t_2-\tau)) - G_i(t_2,\overline{x}(t_2),\phi_{x}(t_2-\tau))   \right) \big| < \displaystyle \frac{L_1 \epsilon_0}{2 \sqrt{n}}
\end{eqnarray}
holds for each $i=1,2,\ldots,n$.

Let us denote by $\eta$ the midpoint of the interval $J$ and set $\alpha=\eta - \xi/2$. There exists an integer $j_0$, $1 \leq j_0 \leq n$, such that
\begin{eqnarray*}
&& \left| G_{j_0}\left(\eta,x(\eta),\phi_{x}(\eta-\tau)\right) - G_{j_0}\left(\eta,\overline{x}(\eta),\phi_{x}(\eta-\tau)\right) \right| \\
&& \geq \displaystyle \frac{1}{\sqrt{n}}  \left\| G\left(\eta,x(\eta),\phi_{x}(\eta-\tau)\right) - G\left(\eta,\overline{x}(\eta),\phi_{x}(\eta-\tau)\right)  \right\|.
\end{eqnarray*}
We obtain by means of the condition $(C2)$ and inequality (\ref{ineqeps0}) that
\begin{eqnarray*}
\left| G_{j_0}\left(\eta,x(\eta),\phi_{x}(\eta-\tau)\right) - G_{j_0}\left(\eta,\overline{x}(\eta),\phi_{x}(\eta-\tau)\right) \right| \geq \displaystyle \frac{L_1}{\sqrt{n}} \left\|x(\eta) - \overline{x}(\eta)\right\| > \frac{L_1 \epsilon_0}{\sqrt{n}}.
\end{eqnarray*}
One can confirm in accordance with inequality (\ref{proofineq1}) that 
\begin{eqnarray} \label{proofineq2}
&& \left| G_{j_0}\left(t,x(t),\phi_{x}(t-\tau)\right) - G_{j_0}\left(t,\overline{x}(t),\phi_{x}(t-\tau)\right) \right| \nonumber \\
&& > \left| G_{j_0}\left(\eta,x(\eta),\phi_{x}(\eta-\tau)\right) - G_{j_0}\left(\eta,\overline{x}(\eta),\phi_{x}(\eta-\tau)\right) \right| - \displaystyle \frac{L_1 \epsilon_0}{2 \sqrt{n}} \nonumber \\
&& > \displaystyle \frac{L_1 \epsilon_0}{2 \sqrt{n}}
\end{eqnarray}
for all $t \in [\alpha, \alpha + \xi]$.

There exist numbers $s_k \in [\alpha, \alpha + \xi]$, $k=1,2,\ldots,n$, such that
\begin{eqnarray*}
&& \displaystyle \Bigg\| \int\limits_{\alpha}^{\alpha+\xi} \left(G(s,x(s),\phi_{x}(s-\tau)) - G(s,\overline{x}(s),\phi_{x}(s-\tau))\right) ds \Bigg\| \\
&& = \displaystyle \Bigg[ \sum\limits_{k=1}^{n} \Bigg( \int\limits_{\alpha}^{\alpha+\xi} \left(G_k(s,x(s),\phi_{x}(s-\tau)) - G_k(s,\overline{x}(s),\phi_{x}(s-\tau))\right) ds \Bigg)^2  \Bigg]^{1/2} \\
&& = \xi \displaystyle \Bigg[ \sum\limits_{k=1}^{n} \left(G_k(s_k,x(s_k),\phi_{x}(s_k-\tau)) - G_k(s_k,\overline{x}(s_k),\phi_{x}(s_k-\tau))\right)^2  \Bigg]^{1/2}.
\end{eqnarray*}
Therefore, the inequality (\ref{proofineq2}) yields
\begin{eqnarray*}
&& \displaystyle \Bigg\| \int\limits_{\alpha}^{\alpha+\xi} \left(G(s,x(s),\phi_{x}(s-\tau)) - G(s,\overline{x}(s),\phi_{x}(s-\tau))\right) ds \Bigg\| \\
&& \geq \xi \left| G_{j_0}(s_{j_0},x(s_{j_0}),\phi_{x}(s_{j_0}-\tau)) - G_{j_0}(s_{j_0},\overline{x}(s_{j_0}),\phi_{x}(s_{j_0}-\tau))  \right| \\
&& > \displaystyle \frac{\xi L_1 \epsilon_0}{2 \sqrt{n}}.
\end{eqnarray*}
Utilizing the equation
\begin{eqnarray*}
 \phi_{x}(t)-\phi_{\overline{x}}(t) & =& \phi_{x}(\alpha)-\phi_{\overline{x}}(\alpha) + \displaystyle  \int\limits_{\alpha}^{t} A \left(\phi_{x}(s)-\phi_{\overline{x}}(s)\right) ds \\
&& +  \displaystyle  \int\limits_{\alpha}^{t} \left(G(s,x(s),\phi_{x}(s-\tau)) - G(s,\overline{x}(s),\phi_{\overline{x}}(s-\tau))\right) ds
\end{eqnarray*}
it can be deduced that
\begin{eqnarray*}
\left\| \phi_{x}(\alpha + \xi)-\phi_{\overline{x}}(\alpha + \xi)\right\| & \geq & \Bigg\|\displaystyle  \int\limits_{\alpha}^{\alpha+\xi} \left(G(s,x(s),\phi_{x}(s-\tau)) - G(s,\overline{x}(s),\phi_{x}(s-\tau))\right) ds \Bigg\| \\
&& - \left\| \phi_{x}(\alpha)-\phi_{\overline{x}}(\alpha) \right\| - \displaystyle \int\limits_{\alpha}^{\alpha+\xi} \left\|A\right\| \left\|  \phi_{x}(s)-\phi_{\overline{x}}(s) \right\| ds \\
&& - \displaystyle  \int\limits_{\alpha}^{\alpha+\xi} \left\| G(s,\overline{x}(s),\phi_{x}(s-\tau)) - G(s,\overline{x}(s),\phi_{\overline{x}}(s-\tau)) \right\| ds\\
&>& \displaystyle\frac{\xi L_1 \epsilon_0}{2\sqrt{n}} - \left(1+\xi \left\|A\right\|+\xi L_3\right) \max_{t\in[\alpha-\tau,\alpha+\xi]}\left\| \phi_{x}(t)-\phi_{\overline{x}}(t)\right\|.
\end{eqnarray*}
Hence,
\begin{eqnarray*}
\displaystyle \max_{t\in[\alpha-\tau,\alpha+\xi]}\left\| \phi_{x}(t)-\phi_{\overline{x}}(t)\right\| > \frac{\xi L_1 \epsilon_0}{2\left(2+\xi \left\|A\right\|+\xi L_3\right)\sqrt{n}}.
\end{eqnarray*}

Suppose that
$
\displaystyle \max_{t\in[\alpha-\tau,\alpha+\xi]}\left\| \phi_{x}(t)-\phi_{\overline{x}}(t)\right\| = \left\| \phi_{x}(\lambda)-\phi_{\overline{x}}(\lambda)\right\|,
$
where $\alpha-\tau \leq \lambda \leq \alpha + \xi$. Let us denote
$$
\epsilon_1=\frac{\xi L_1 \epsilon_0}{4\left(2+\xi \left\|A\right\|+\xi L_3\right)\sqrt{n}}
$$
and
$$
\overline{\Delta}=\frac{\xi L_1 \epsilon_0}{4\left(\left\|A\right\|M_0+M_G\right)\left(2+\xi \left\|A\right\|+\xi L_3\right)\sqrt{n}}.
$$
For $t \in \widetilde{J}$, where $\widetilde{J}=\left[ \lambda-\overline{\Delta}/2,\lambda+\overline{\Delta}/2 \right]$, we have that
\begin{eqnarray*}
\displaystyle \left\| \phi_{x}(t)-\phi_{\overline{x}}(t)\right\| & \geq & \left\| \phi_{x}(\lambda)-\phi_{\overline{x}}(\lambda)\right\| - \displaystyle \Bigg| \int\limits_{\lambda}^{t} \left\|A\right\|  \left\| \phi_{x}(s)-\phi_{\overline{x}}(s)\right\| ds \Bigg| \\
&& -  \Bigg| \displaystyle  \int\limits_{\lambda}^{t}  \left\|G(s,x(s),\phi_{x}(s-\tau)) - G(s,\overline{x}(s),\phi_{\overline{x}}(s-\tau)) \right\| ds  \Bigg| \\
&& > \frac{\xi L_1 \epsilon_0}{2\left(2+\xi \left\|A\right\|+\xi L_3\right)\sqrt{n}} - \overline{\Delta} \left( \left\|A\right\|M_0+M_G\right).  
\end{eqnarray*}
Thus, $\displaystyle \left\| \phi_{x}(t)-\phi_{\overline{x}}(t)\right\|>\epsilon_1$ for all $t\in\widetilde{J}$.
Consequently, system (\ref{system2}) replicates the sensitivity of system (\ref{system1}). $\square$

It is worth noting that if an autonomous system is utilized instead of the non-autonomous system (\ref{system1}) as the generator, then the result obtained in Theorem \ref{delaymaintheorem1} is also valid with the counterpart of condition $(C1)$.
This is illustrated in the first example provided in Section \ref{secexamples}.

The next section is devoted to the replication of period-doubling cascade.

\section{Replication of Period-Doubling Cascade} \label{pdcsection} 

Let us consider the system
\begin{eqnarray} \label{pdcsystem1}
x'(t)=H(t,x(t),\mu),
\end{eqnarray}
where $\mu$ is a real parameter and the function $H:\mathbb R \times \mathbb R^m\times\mathbb R\to \mathbb R^m$, which is continuous in all of its arguments, satisfies the equation $H(t+T,x,\mu)=H(t,x,\mu)$ for all $t\in \mathbb R$, $x \in \mathbb R^m$, and $\mu \in \mathbb R$. We suppose that there exists a finite value $\mu_{\infty}$ of the parameter $\mu$ such that the function $F(t,x)$ on the right-hand side of system (\ref{system1}) is equal to $H(t,x,\mu_{\infty})$.

System (\ref{system1}) is said to admit a period-doubling cascade \cite{Feigenbaum80,Sander11,Alligood96} if there exists a sequence $\left\{\mu_j\right\}$, $\mu_j \to \mu_{\infty}$ as $j\to \infty$, of period-doubling bifurcation values such that system (\ref{pdcsystem1}) undergoes a period-doubling bifurcation as the parameter $\mu$ increases or decreases through each $\mu_j$, i.e., for each $j \in \mathbb N$ a new stable periodic solution with period $p_0 2^j T$ appears in the dynamics of (\ref{pdcsystem1}) for some positive integer $p_0$, and the preceding $p_0 2^{j-1} T$-periodic solution loses its stability. Therefore, at the parameter value $\mu=\mu_{\infty}$ there exist infinitely many unstable periodic solutions of system (\ref{pdcsystem1}), and hence of system (\ref{system1}), all lying in a bounded region.

We say that system (\ref{system2}) replicates the period-doubling cascade of system (\ref{system1}) if for each periodic solution $x \in \mathscr{A}$ of (\ref{system1}) system (\ref{system2}) admits a periodic solution with the same period.

The one-to-one correspondence between the periodic solutions of systems (\ref{system1}) and (\ref{system2}) is mentioned in the following lemma.
\begin{lemma} \label{pdclemma}
Suppose that the conditions $(C2)$, $(C4)$, $(C5)$ hold and $\omega-K L_3>0$. Then $x\in\mathscr{A}$ is a $k_0 T$-periodic solution of the generator system (\ref{system1}) for some positive integer $k_0$ if and only if the bounded solution $\phi_{x} \in\mathscr{B}$ of the replicator system (\ref{system2}) is $k_0 T$-periodic.
\end{lemma}
\noindent \textbf{Proof.} First suppose that $x \in \mathscr{A}$ is a $k_0 T$-periodic solution of the generator system (\ref{system1}).
Using the integral equation (\ref{bddsolnrelation}) we obtain that
\begin{eqnarray*}
 \left\|\phi_{x}(t+k_0T)-\phi_{x}(t)\right\|  & \leq & \displaystyle \int\limits_{-\infty}^{t} \left\|e^{A(t-s)}\right\|  \left\|G(s,x(s),\phi_{x}(s+k_0 T-\tau)) - G(s,x(s),\phi_{x}(s-\tau))  \right\| ds \\
& \leq & \displaystyle \frac{KL_3}{\omega} \sup_{t\in\mathbb R} \left\|\phi_{x}(t+k_0T)-\phi_{x}(t)\right\|.
\end{eqnarray*}
The last inequality implies that $\displaystyle\sup_{t\in\mathbb R} \left\|\phi_{x}(t+k_0T)-\phi_{x}(t)\right\|=0$. Thus, $\phi_{x}(t)$ is $k_0T$-periodic.

Conversely, let us assume that $\phi_{x}\in \mathscr{B}$ is $k_0T$-periodic. Then we have $$G(t,x(t),\phi_{x}(t-\tau)) = G(t,x(t+k_0T),\phi_{x}(t-\tau))$$ for all $t \in \mathbb R$. Using the last equation and condition $(C2)$, one can confirm that $x \in \mathscr{A}$ is $k_0 T$-periodic. $\square$

It is worth noting that if $x \in \mathscr{A}$ is an unstable periodic solution of system (\ref{system1}), then the periodic solution $(x,\phi_{x}) \in \mathscr{A} \times \mathscr{B}$ of the coupled system (\ref{system1})-(\ref{system2}) is also unstable. 

The following theorem can be proved by using Lemma \ref{pdclemma}. 

\begin{theorem} \label{pdctheorem} Suppose that the conditions $(C1)-(C6)$ hold. If system (\ref{system1}) admits a period-doubling cascade, then system (\ref{system2}) replicates the period-doubling cascade of (\ref{system1}).
\end{theorem}

The result of Theorem \ref{pdctheorem} is also valid in the case that the generator system is an autonomous one with the counterpart of condition $(C1)$ and the periods of its periodic solutions appearing in the cascade and the number $T$ satisfying (\ref{periodofG}) are commensurable.

A corollary of Theorem \ref{pdctheorem} is as follows.

\begin{corollary} \label{pdccorollary} Suppose that the conditions $(C1)-(C6)$ hold. If system (\ref{system1}) admits a period-doubling cascade, then the same is true for the coupled system (\ref{system1})-(\ref{system2}).
\end{corollary}

It is worth noting that the coupled system (\ref{system1})-(\ref{system2}) possesses exactly the same sequence of period-doubling bifurcation values with the generator system (\ref{system1}) under the conditions of Theorem \ref{pdctheorem}. For that reason the Feigenbaum universality \cite{Feigenbaum80} holds also for the coupled system (\ref{system1})-(\ref{system2}) provided that it is valid for (\ref{system1}).

\section{Examples}\label{secexamples}

Two illustrative examples that support the theoretical results are provided in this section. The replication of sensitivity is discussed in the first example, and the second one is concerned with replication of period-doubling route to chaos.

\subsection{Example 1}

Let us consider the Lorenz system \cite{Lorenz63,Sparrow82}
\begin{eqnarray} \label{examplesystem1}
&& x'_1(t)=-10x_1(t) + 10x_2(t) \nonumber \\
&& x'_2(t)= -x_1(t) x_3(t) +28 x_1(t) -x_2(t) \\
&& x'_3(t)= x_1(t) x_2(t) - \frac{8}{3} x_3(t). \nonumber  
\end{eqnarray} 
It was demonstrated by Tucker \cite{Tucker99} that system (\ref{examplesystem1}) admits a chaotic attractor. 

In this example, we use the Lorenz system (\ref{examplesystem1}) as the generator, and as the replicator we take into account the system 
\begin{eqnarray} \label{examplesystem2}
&& y'_1(t)= -4.5 y_1(t) + 0.3\tanh(y_2(t-0.25)) + 1.7 x_1(t) \nonumber \\
&& y'_2(t)= -2.8 y_2(t) + 0.4\sin(y_1(t-0.25)) - 1.5 x_2(t) \\
&& y'_3(t)= -3.6 y_3(t)+ 0.8 x_3(t) +0.1\cos t, \nonumber  
\end{eqnarray} 
where $(x_1(t), x_2(t), x_3(t))$ is a solution of system (\ref{examplesystem1}).

System (\ref{examplesystem2}) is in the form of (\ref{system2}) with 
$$A=\textrm{diag}(-4.5, -2.8, -3.6),$$ 
$$G(t,x_1,x_2,x_3,y_1,y_2,y_3)= (0.3\tanh y_2 + 1.7 x_1,0.4\sin y_1 - 1.5 x_2, 0.8 x_3 +0.1\cos t),$$ and $\tau=0.25$. 

The conditions of Theorem \ref{delaymaintheorem1} are satisfied for the coupled system (\ref{examplesystem1})-(\ref{examplesystem2}) with $K=1$, $\omega=2.8$, $L_1=0.8$, $L_2=1.7$, $L_3=0.4$, and accordingly, system (\ref{examplesystem2}) replicates the sensitivity of the Lorenz system (\ref{examplesystem1}).

\begin{figure}[ht!] 
\centering
\includegraphics[width=10.5cm]{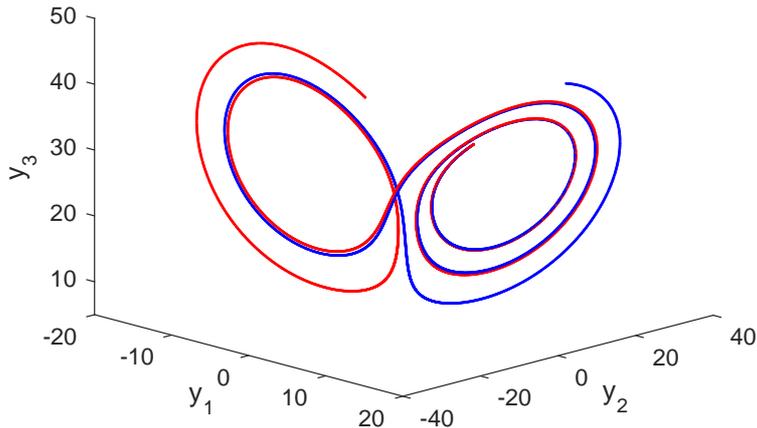}
\caption{Replication of sensitivity by system (\ref{examplesystem2}). The figure supports the result of Theorem \ref{delaymaintheorem1} such that two trajectories of the replicator system (\ref{examplesystem2}) which are nearby on the interval $[-0.25,0]$ eventually diverge.}
\label{delaypdcfig1}
\end{figure}

In order to illustrate the replication of sensitivity, we depict in Figure \ref{delaypdcfig1} the trajectories of two initially nearby solutions of system (\ref{examplesystem2}) in which initially nearby solutions of (\ref{examplesystem1}) that eventually diverge are utilized. Let us consider the constant functions $u_1(t)= -1.06$, $u_2(t)= 3.28$, $u_3(t)= 4.36$, $v_1(t)= -1.04$, $v_2(t)= 3.26$, and $v_3(t)= 4.37$. Using the solution $(x_1(t),x_2(t),x_3(t))$ of (\ref{examplesystem1}) with $x_1(0)=7.93$, $x_2(0)=2.41$, $x_3(0)=33.05$ in (\ref{examplesystem2}), we obtain the trajectory shown in blue corresponding to the initial data $y_1(t)=u_1(t)$, $y_2(t)=u_2(t)$, $y_3(t)=u_3(t)$, $t \in [-0.25,0]$. On the other hand, the trajectory in red represents the solution of (\ref{examplesystem2}) corresponding to $y_1(t)=v_1(t)$, $y_2(t)=v_2(t)$, $y_3(t)=v_3(t)$, $t \in [-0.25,0]$, when the solution $(x_1(t),x_2(t),x_3(t))$ of (\ref{examplesystem1}) with $x_1(0)=7.97$, $x_2(0)=2.36$, $x_3(0)=33.09$ is utilized in (\ref{examplesystem2}). The time interval $[0, 3.18]$ is used in the simulation. Figure \ref{delaypdcfig1} confirms the result of Theorem \ref{delaymaintheorem1} such that the trajectories in blue and red eventually diverge even if they are nearby on the interval $[-0.25,0]$.

\begin{figure}[ht!] 
\centering
\includegraphics[width=15.0cm]{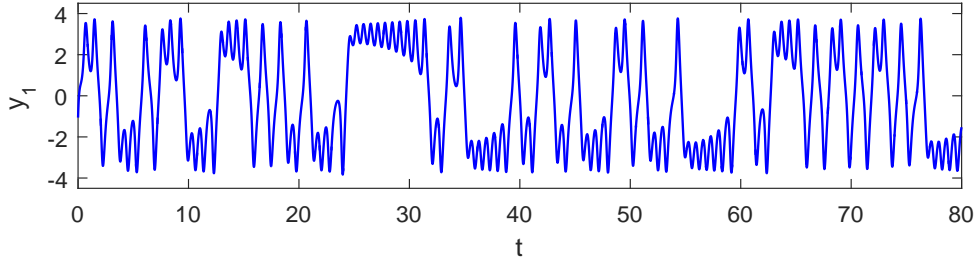}
\caption{Time series of the $y_1$-coordinate of the coupled system (\ref{examplesystem1})-(\ref{examplesystem2}). The figure reveals the chaotic behavior in the dynamics of the replicator system (\ref{examplesystem2}).}
\label{delaypdcfig2}
\end{figure}

Next, to demonstrate the chaotic behavior of system (\ref{examplesystem2}), using the solution $(x_1(t),x_2(t),x_3(t))$ of (\ref{examplesystem1}) with $x_1(0)=7.93$, $x_2(0)=2.41$, $x_3(0)=33.05$ one more time, we represent in Figure \ref{delaypdcfig2} the time series of the $y_1$-coordinate of (\ref{examplesystem2}) corresponding to the initial data $y_1(t)=u_1(t)$, $y_2(t)=u_2(t)$, $y_3(t)=u_3(t)$ for $t \in [-0.25,0]$. The irregularity seen in Figure \ref{delaypdcfig2} manifests the replication of chaos.

\subsection{Example 2}

Let us take into account the Duffing equation
\begin{eqnarray} \label{examplesystem3}
x''(t)+0.3x'(t)+x^3(t)=\mu \cos t,
\end{eqnarray} 
where $\mu$ is a parameter.
It was shown by Sato et al. \cite{Sato83} that equation (\ref{examplesystem3}) displays period-doubling bifurcations and leads to chaos at $\mu=\mu_{\infty} \equiv 40$.

Using the new variables $x_1(t)=x(t)$ and $x_2(t)=x'(t)$, equation (\ref{examplesystem3}) can be rewritten as the system
\begin{eqnarray} \label{examplesystem4} 
&& x'_1(t)=x_2(t) \nonumber \\
&& x'_2(t)=-0.3x_2(t)-x_1^3(t)+ \mu \cos t.
\end{eqnarray} 

One can confirm that the chaotic attractor of system (\ref{examplesystem4}) with $\mu=\mu_{\infty}$ takes place inside the compact region
$$
\Lambda= \left\{(x_1,x_2)\in \mathbb R^2 :~ \left| x_1 \right| \leq 5.5, ~ \left| x_2 \right| \leq 14  \right\}.
$$ 

Next, we consider the system with delay
\begin{eqnarray} \label{examplesystem5} 
&& y'_1(t)= -2y_1(t)+y_2(t) + 1.3 x_1(t) -0.06 x^2_1(t)+ 0.1 \sin t \nonumber \\
&& y'_2(t)= -0.5 y_1(t)-3 y_2(t)+0.14\arctan(y_1(t-0.15))+0.9 x_2(t),
\end{eqnarray} 
where $(x_1(t),x_2(t))$ is a solution of system (\ref{examplesystem4}). The system (\ref{examplesystem4})-(\ref{examplesystem5}) is a unidirectionally coupled one in which (\ref{examplesystem4}) is the generator and (\ref{examplesystem5}) is the replicator.

Systems (\ref{examplesystem4}) and (\ref{examplesystem5}) are respectively in the forms of (\ref{system1}) and (\ref{system2}), where $$F(t,x_1,x_2)=\left( x_2,  -0.3x_2-x_1^3+ \mu \cos t \right),$$ $$A=\begin{pmatrix} -2 & 1 \\ -0.5 & -3 \end{pmatrix},$$ $$G(t,x_1,x_2,y_1,y_2)=\left( 1.3 x_1 -0.06 x^2_1+ 0.1 \sin t , 0.14\arctan y_1+0.9 x_2 \right),$$ and $\tau=0.15$. The eigenvalues of the matrix $A$ are $\displaystyle -\frac{5}{2}+\frac{1}{2}i$ and $\displaystyle -\frac{5}{2}-\frac{1}{2}i$.
Let us denote $$P=\begin{pmatrix} 0 & 1 \\ 0.5 & -0.5 \end{pmatrix}.$$ Using the equation
$$e^{At}=e^{-5t/2} P \begin{pmatrix}  \cos \left(\frac{t}{2}\right) & -\sin\left(\frac{t}{2}\right) \\ \sin\left(\frac{t}{2}\right) & \cos\left(\frac{t}{2}\right) \end{pmatrix} P^{-1},$$
it can be verified that $\left\|e^{At}\right\| \leq K e^{-\omega t}$ for all $t \geq 0$, where $K=\left\|P\right\|\left\|P^{-1}\right\| \approx 2.618034$ and $\omega=2.5$.

The conditions $(C1)-(C6)$ are valid for systems (\ref{examplesystem4}) and (\ref{examplesystem5}) with $L_F=90.76$, $L_1=0.452549$, $L_2=2.156757$, $L_3=0.14$, and $M_G=15.701095.$ According to our theoretical results, system (\ref{examplesystem5}) replicates the period-doubling cascade of system (\ref{examplesystem4}), and the coupled system (\ref{examplesystem4})-(\ref{examplesystem5}) is chaotic at the parameter value $\mu=\mu_{\infty}$.

\begin{figure}[ht!] 
\centering
\includegraphics[width=16.0cm]{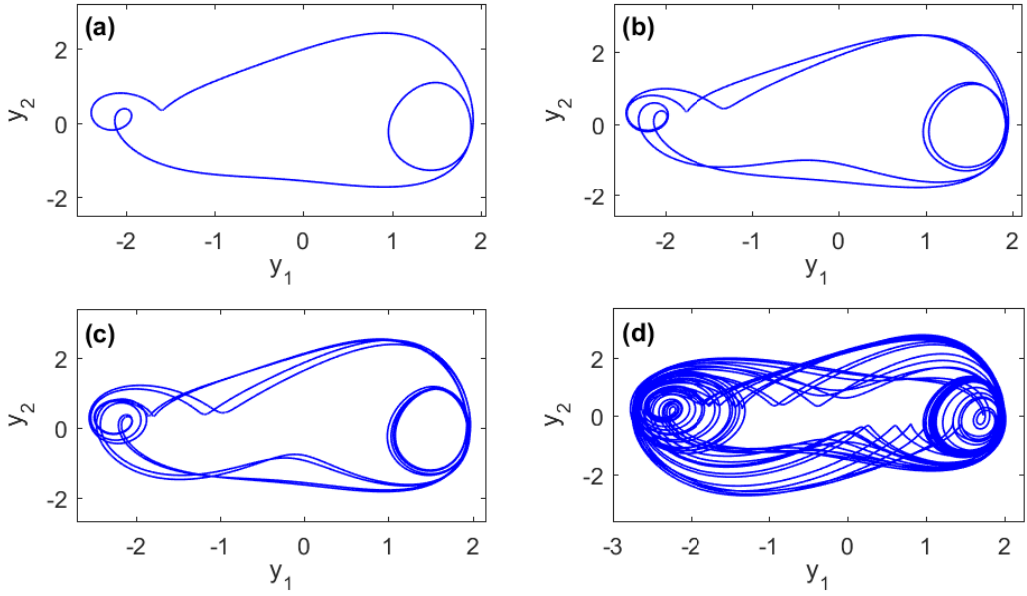}
\caption{Projections of periodic and irregular orbits of the coupled system (\ref{examplesystem4})-(\ref{examplesystem5}) on the $y_1-y_2$ plane. (a) Period-$1$ orbit. (b) Period-$2$ orbit. (c) Period-$4$ orbit. (d) Irregular orbit. The values $31.7$, $34.3$, $36.1$, and $40$ of the parameter $\mu$ are respectively used in (a), (b), (c), and (d). The figure reveals the replication of period-doubling route to chaos.}
\label{delaypdcfig3}
\end{figure}

\begin{figure}[ht!] 
\centering
\includegraphics[width=15.8cm]{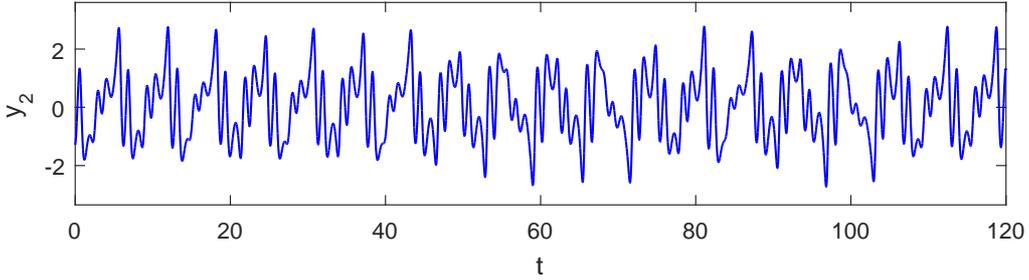}
\caption{Irregular behavior in the dynamics of the replicator system (\ref{examplesystem5}). The figure shows the time series of the $y_2$-coordinate of the coupled system (\ref{examplesystem4})-(\ref{examplesystem5}) with $\mu=40$ corresponding to the initial data $x_1(t)=u_1(t)$, $x_2(t)=u_2(t)$, $y_1(t)=u_3(t)$, $y_2(t)=u_4(t)$ for $-0.15 \leq t \leq 0$, where $u_1(t)=1.26$, $u_2(t)=-2.21$, $u_3(t)=1.36$, and $u_4(t)=-1.29$.}
\label{delaypdcfig4}
\end{figure}

Figure \ref{delaypdcfig3} depicts the projections of periodic and irregular orbits of the coupled system (\ref{examplesystem4})-(\ref{examplesystem5}) on the $y_1-y_2$ plane. The projections of period-$1$, period-$2$, and period-$3$ orbits are shown in Figure \ref{delaypdcfig3}, (a), (b), and (c), respectively. The values $31.7$, $34.3$, and $36.1$ of the parameter $\mu$ are respectively used in Figure \ref{delaypdcfig3}, (a), (b), and (c). Figure \ref{delaypdcfig3}, (d), on the other hand, represents the projection of the irregular orbit for $\mu=40$ corresponding to the initial data $x_1(t)=u_1(t)$, $x_2(t)=u_2(t)$, $y_1(t)=u_3(t)$, $y_2(t)=u_4(t)$ for $-0.15 \leq t \leq 0$, where $u_1(t)=1.26$, $u_2(t)=-2.21$, $u_3(t)=1.36$, and $u_4(t)=-1.29$ are constant functions. The time series of the $y_2$-coordinate of the solution of the coupled system (\ref{examplesystem4})-(\ref{examplesystem5}) corresponding to the same initial data and the same value of $\mu$ that are utilized in Figure \ref{delaypdcfig3}, (d) is shown in Figure \ref{delaypdcfig4}. Figures \ref{delaypdcfig3} and \ref{delaypdcfig4} manifest that system (\ref{examplesystem5}) replicates the period-doubling cascade of (\ref{examplesystem4}).

\newpage

\section{Conclusions} \label{secconc}

This paper is devoted to replication of chaos for unidirectionally coupled systems in which the replicator is a system with delay. It is rigorously proved that the replicator exhibits dynamics similar to the one of the generator system, which is the source of chaotic motions. The results are based on the replication of sensitivity and the existence of infinitely many unstable periodic solutions in a compact region. Due to the presence of delay, a novel definition as well as a more complicated proof for the replication of sensitivity are provided compared to the paper \cite{Fen13}. Using the technique presented in this paper it is possible to obtain high dimensional systems with delay which possess chaotic motions. The obtained theoretical results may be applied to various fields such as neural networks, secure communication, robotics, economics, and lasers in which dynamics are described through differential equations with delay \cite{Lakshmanan18}-\cite{Erneux19}.

\end{document}